\documentclass[twocolumn,showpacs,preprintnumbers,amsmath,amssymb]{revtex4}

\usepackage{graphicx}

\begin{document}


\title{Photoconductance Quantization in a Single-Photon Detector }
\author{Hideo Kosaka$^{1\ast }$, Deepak S. Rao$^{1}$, Hans D. Robinson$^{1}$,
Prabhakar Bandaru$^{1}$, Toshitsugu Sakamoto$^{2}$, Eli Yablonovitch$^{1}$}
\address{$^{1}$ Electrical Engineering Department, University of California
Los Angeles, Los Angeles, CA, 90095-1594\\
$^{2}$ Fundamental Research Laboratories, NEC Corporation, 34
Miyukigaoka, Tsukuba, Ibaraki 305-8501, Japan\\
$^{\ast }$ On leave from Fundamental Research Laboratories, NEC Corporation.}

\begin{abstract}
We have made a single-photon detector that relies on
photoconductive gain in a narrow electron channel in an
AlGaAs/GaAs 2-dimensional electron gas. Given that the electron
channel is 1-dimensional, the photo-induced conductance has
plateaus at multiples of the quantum conductance 2e$^{2}$/h.
Super-imposed on these broad conductance plateaus are many sharp,
small, conductance steps associated with single-photon absorption
events that produce individual photo-carriers. This type of
photoconductive detector could measure a single photon, while
safely storing and protecting the spin degree of freedom of its
photo-carrier. This function is valuable for a quantum repeater
that would allow very long distance teleportation of quantum
information.
\end{abstract}

\pacs{72.20.-i, 72.40.+w, 73.23.-b, 73.63.-b, 78.67.-n}
\maketitle

The detection of individual photons has become common in the last
few decades with the proliferation of avalanche photodetectors,
\cite {Yoshizawa01} and negative electron affinity
\cite{Antypas74} photocathode photomultipliers. However, it is
becoming desirable now to transmit something more sophisticated
than single-photon states. Namely, we want to distribute quantum
entanglement information over long distances to enable
teleportation \cite{Bennett93} and other forms of advanced
telecommunication. For this, single-photon sensitivity is not
enough. New types of single-photon photodetectors are needed,
which preserve the spin information of the photo-carrier.

New selection rules have been identified in the III-V semiconductor
photodetectors, that permit the transfer \cite{Vrijen01} of photon
polarization information directly to photo-electron spin. Unfortunately, the
avalanche multiplication process is relatively violent, of necessity, and it
destroys any spin information that might have been present on the original
photo-electron. New forms of single-photon detection have been developed
recently, that are gentle enough to preserve the photo-electron spin state,
while providing the gain needed to detect a single photon. For example
photoconductive gain can provide for single-photon detection.

Shields et al.\cite{Shields00} recently demonstrated single-photon
sensitivity in a two-dimensional (2D) electron channel controlled
by photoelectric charge trapped on adjacent In$_{x}$Ga$_{1-x}$As
precipitates in AlGaAs. In photoconductive gain, electric charges
become trapped for long periods, and while trapped, their electric
charge can influence the channel conductivity. Over time, vast
amounts of electric charge can be transferred in the channel
current, far more than the original trapped charge, creating a
huge photoconductive gain. Trapping of an individual photo-carrier
produces a step function in current that can persist for long
periods. In AlGaAs alloys at low temperature this is called
persistent photoconductivity, and the current can last for weeks.
An analogous electrical transport sensitivity enhancement was
spectacularly used to detect individual far-infrared photons by
Komiyama et al.\cite{Komiyama00}

\begin{figure}
\includegraphics[width=8.1cm]{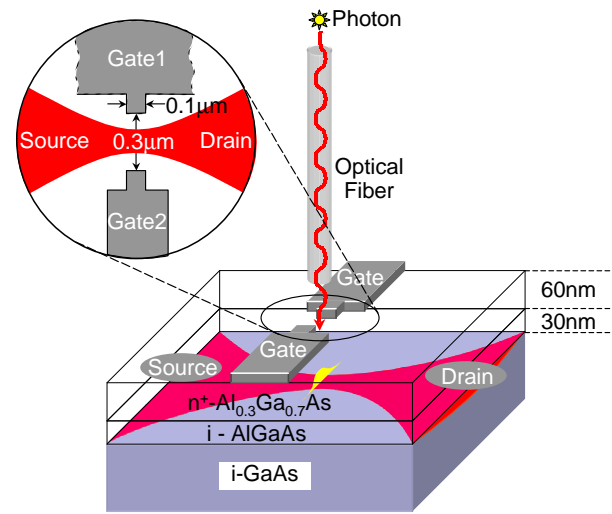}
\caption{\label{fig1} Schematic diagram of the experimental setup.
A nearly pinched off point contact channel in a 2D electron gas at
an AlGaAs/GaAs interface can act as a sensitive photodetector. The
source/drain channel conductance can sense individual trapped
photo-carriers, whose effect is multiplied by the photoconductive
gain mechanism.}
\end{figure}

In this report, we demonstrate that an ordinary, conventional, Al$_{y}$Ga$%
_{1-y}$As/GaAs modulation-doped field effect transistor, in a 2D electron
gas quantum point contact configuration, can already detect single-photon
events. In$_{x}$Ga$_{1-x}$As precipitates in GaAs are unnecessary, as the
ordinary defect centers that are already present in Al$_{y}$Ga$_{1-y}$%
As/GaAs can trap photo-carriers successfully. Furthermore, the trapped
photocharge has the same effect as a positive external gate in un-pinching
the channel, leading to conductance quantization steps. Thus we are in the
interesting situation where individual single-photon-conductivity steps are
superimposed on the 1D channel conductance quantization steps associated
with a variable width point contact channel. As an aside, this may also be
the first observation of photo-induced 1D channel conductance quantization.

\begin{figure}
\includegraphics[width=8.1cm]{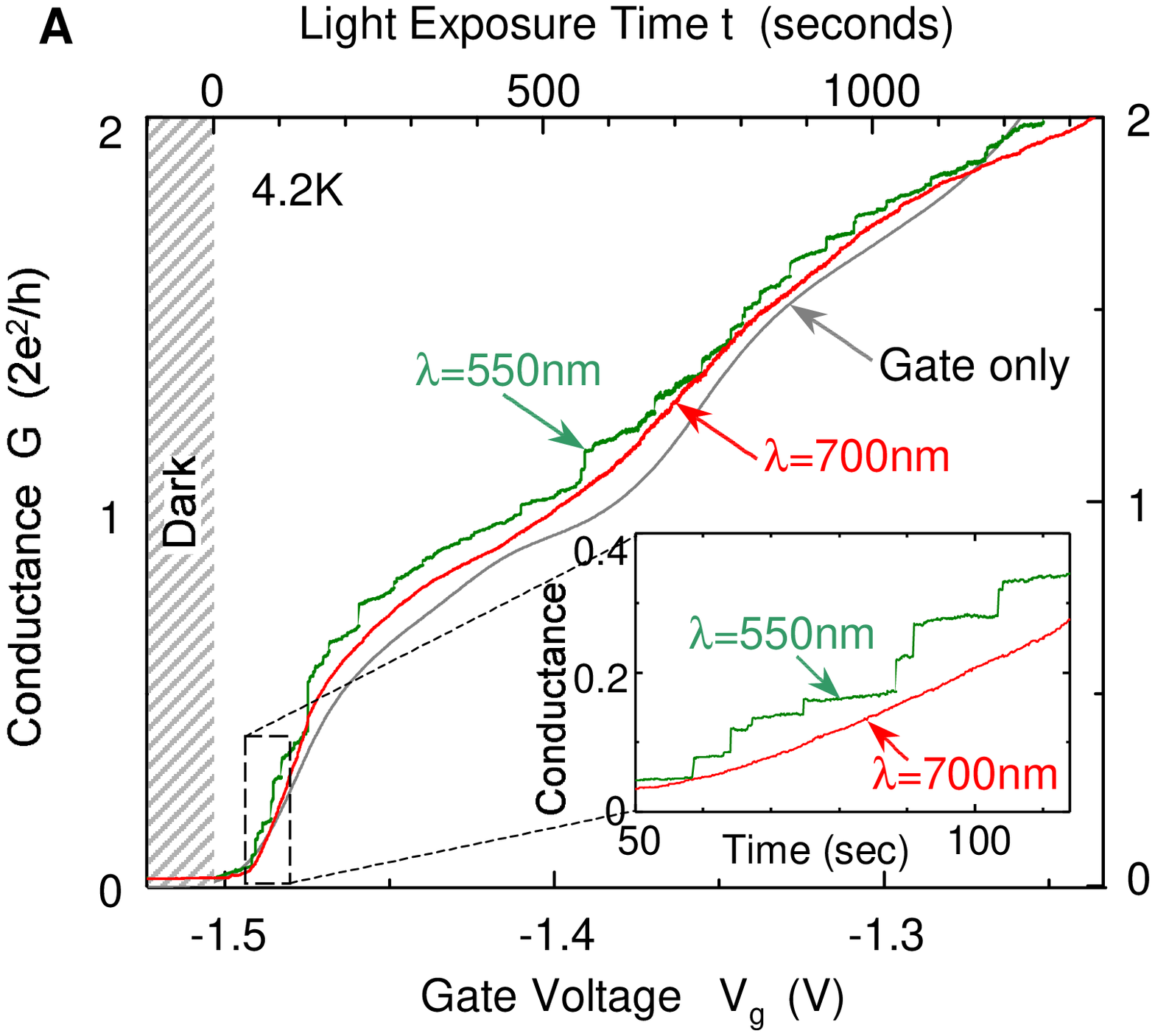}
\includegraphics[width=8.1cm]{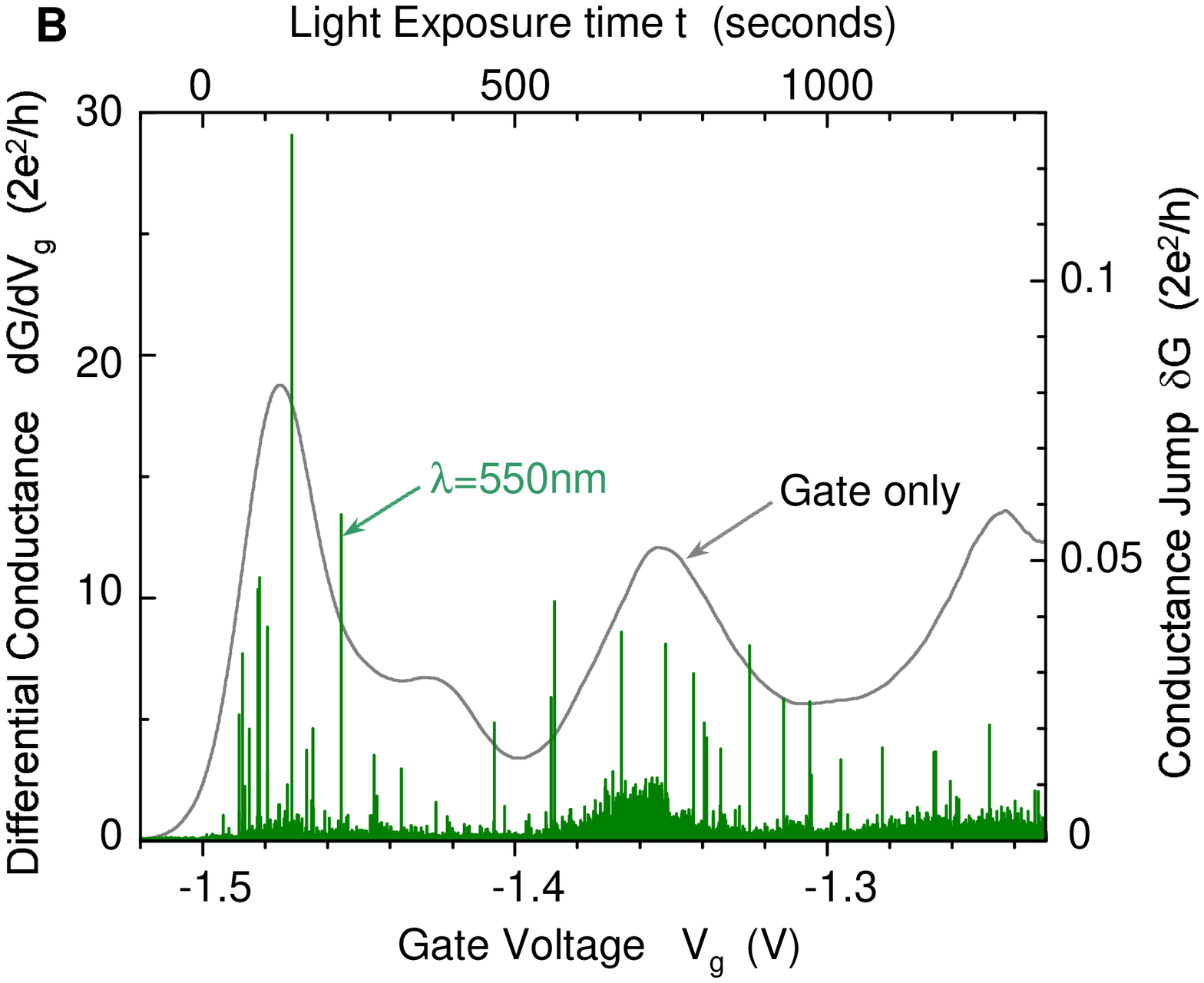}
\caption{\label{fig2} (A) The source/drain channel conductance
induced by a modulation of a gate voltage, and by light exposure
for a period of time. The conductance quantization plateaus at
multiples of 2e$^{2}$/h are practically identical. With light at
$\lambda $ = 550 nm, there is the additional feature of individual
small conductance steps associated with single photons. (B)
Differential conductance dG/dV$_{g}$ versus gate voltage, and
individual photon conductance jump height versus light exposure.
The single-photon conductance jumps are larger at those bias
conditions where the channel conductance is more sensitive to gate
voltage.}
\end{figure}

The sample used for these experiments is a modulation-doped heterostructure
with a quantum point contact defined by a pair of split gates on Al$_{0.3}$Ga%
$_{0.7}$As/GaAs, as illustrated in Fig. 1. All layers are grown by molecular
beam epitaxy on semi-insulating GaAs, consisting of a nominally undoped GaAs
buffer layer, an i-Al$_{0.3}$Ga$_{0.7} $As spacer layer 30 nm thick, a
Si-doped (1$\times $10$^{18}$cm$^{-3}$) n-Al$_{0.3}$Ga$_{0.7}$As layer 60 nm
thick, and a Si-doped GaAs cap layer, 5 nm. The 2D electron gas in the
heterointerface has a carrier density of 3.3$\times $10$^{11}$ cm$^{-2}$, a
mobility of 1.1$\times $10$^{6}$ cm$^{2}$/V s and a Fermi energy (E$_{F}$)
of 1.8 meV. The Ti/Au split gate, of lithographic length 0.1 $\mu $m, and
spacing 0.3 $\mu $m, is fabricated on the heterostructure using
electron-beam lithography and electron-gun evaporation. The 1D channel is
formed as a line by depletion in the 2DEG between the two gates.

The sample is illuminated by monochromatic light through a large-core glass
fiber, that is carefully shielded to block any photons from the outer
jacket. The light is created by a tungsten lamp and then filtered by a
monochromator, a long-pass filter passing wavelengths $\lambda $
\mbox{$>$}%
530 nm, and a 20 dB neutral density filter. The optical power at the end of
the fiber measured by a Si detector is $\approx $ 9 pico-Watts. The area of
illumination at the device is about 1 mm in diameter due to end-fire
coupling from the fiber. Given the small device area 3$\times $10$^{-10}$ cm$%
^{2}$ defined by the gates, we estimate the actual light power in the active
area to be 7$\times $10$^{-9}$ times smaller. Thus the incident photon flux
is estimated to be 0.1 photon per second on the effective device area.

The source/drain current is measured at a constant voltage drop (V$_{SD}$)
of 0.5 mV, at a temperature of 4.2 K. Figure 2A shows the corresponding
source/drain conductance, as a function of either gate voltage, or of the
light exposure time. As the channel is opened up, a series of electron
wave-guided modes successively contribute conductance steps,\cite{van
Wees88,Wharam88} in units of the conductance quantum, 2e$^{2}$/h $\approx $
1/13,000 $\Omega $, where e is the electronic charge, the factor 2 accounts
for spin, h is Planck's constant, and $\Omega $ is Ohms. The first two steps
are shown in Fig. 2A, and their sharpness is consistent with the
temperature, 4.2 K. In addition there is a well-known \cite{Thomas96}
shoulder at conductance 0.7 (2e$^{2}$/h) thought to be associated with
electron spin exchange interaction \cite{Wang98} effects.

What is remarkable about Fig. 2A is that there are two different physical
phenomena, producing almost identical source/drain conductance on the
vertical axis. The curve labeled ``gate only'' shows that positive gate
voltage, above the -1.5 Volt gate threshold, opens up the electron channels
producing conductance steps. Likewise, exposure to a weak light source of
wavelength $\lambda $ = 700 nm, at a fixed bias voltage produces trapped
positive charge that also opens up the electron waveguide channels,
producing exactly the same conductance steps. In fact the processes are the
same. In either case, positive net charge opens up the source/drain electron
current channels, leading to the observed electron conductance steps.

An enduring, photo-induced, increase in conductivity has been well known
\cite{Nelson77,Kastalsky84,Wei84} in III-V semiconductors, and is called
persistent photoconductivity. At temperatures much lower than 100K the net
positive trapped charge, is known to persist for weeks. The photo-exposure
begins at time t = 0 in Fig. 2A, to the right of the crosshatched dark
region where the conductivity begins as a constant. If the photo-exposure is
prematurely terminated, the conductance becomes constant again in Fig. 2A,
persisting at the new value for weeks.

Photoconductivity in an electron channel requires fixed positive charge from
trapped photo-holes. The hole trapping centers can be either neutral donors d%
$^{0}$, that become ionized undergoing the transition d$^{0}$ + h$^{+}$ $%
\rightarrow $ d$^{+}$, or they can be DX$^{-}$ centers \cite
{Lang77,Lang79,Chadi88,Chadi89,Linke98} that become neutralized by hole
capture,\cite{Brunthaler89} DX$^{-}$ + h$^{+}$ $\rightarrow $ d$^{0}$. In
any case, the net trapped positive charge has the same effect on the
source/drain electron channel as positive gate increments do. The only
difference in Fig. 2A is that the horizontal axis at the top measures the
net positive charge in terms of optical exposure time from a weak $\lambda $
= 700 nm beam, and the horizontal axis at the bottom measures positive
increase of gate voltage. The same quantum conductance plateaus are produced
in either case.

When the photon wavelength is reduced to $\lambda $ = 550 nm, well beyond
the Al$_{0.3}$Ga$_{0.7}$As bandgap, an additional phenomenon appears, that
is plotted in Fig. 2A, as the curve labeled ``$\lambda $ = 550 nm''. That
curve still follows the overall shape of the quantized conductance steps,
but the curve itself consists of many smaller steps that, in aggregate,
produce the quantized conductance shape, including the 0.7 (2e$^{2}$/h)
feature. The smaller steps, we attribute to absorption of individual
photons, and the corresponding capture of a single photo-hole. Since the
traps are at variable distances from the source/drain channel, each photon
produces a different step height. This is different from the Millikan
oil-drop experiment, but similar to the observations by Shields et al.,\cite
{Shields00} who trapped photo-holes on InGaAs islands.

\begin{figure}
\includegraphics[width=8.1cm]{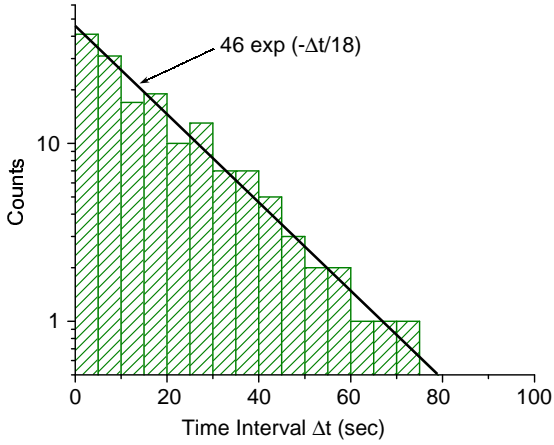}
\caption{\label{fig3} Statistics of the time intervals between
single-photon detection events. The exponential decay is
consistent with un-squeezed photon statistics.}
\end{figure}

The photon steps for $\lambda $ = 550 nm in Figs. 2A seem to be taller where
the conductance curve is steeper, due to greater sensitivity to
electrostatic charge changes when dG/dV$_{g}$ is larger. This point is
illustrated in Fig. 2B, that plots: (1) dG/dV$_{g}$ versus gate voltage on
the left and bottom axes, and on the same graph, (2) $\delta $G,
single-photon step height versus photon exposure time on the right and top
axes, respectively.

The $\lambda $ = 550 nm ``curve'' in Fig. 2A is seen in Fig. 2B to consist
of about 70 individual photon steps. We can test for proper photon
statistics by plotting a histogram of the time intervals between photon
events, as illustrated in Fig. 3. The intervals should fall on a decaying
exponential for random photon events, as is appropriate for un-squeezed
photon statistics, with the average interval between photon events being 18
seconds in this case. This low photon detection rate is consistent with an
active area of 3$\times $10$^{-10}$ cm$^{2}$, and a quantum efficiency of
around 30\%. The monotonic upward conductance changes are to be
distinguished from ``random telegraph signal'', \cite{Ralls84} that
fluctuates in either direction.

\begin{figure}
\includegraphics[width=8.1cm]{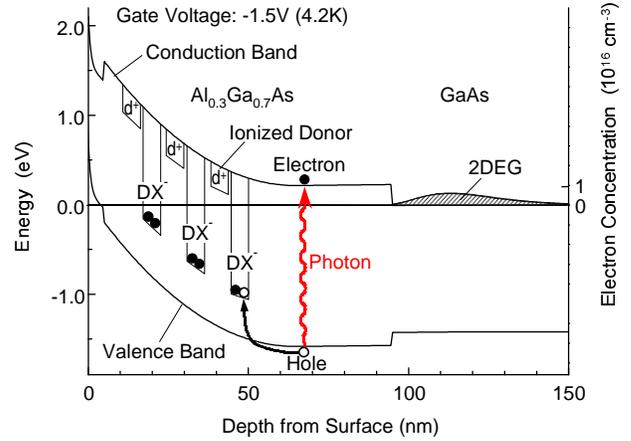}
\caption{\label{fig4} A model for the single-photon detection,
carrier capture, and the photoconductive gain mechanism. A
photo-hole is trapped at either a DX$^{-}$ center or a neutral
donor d$^{0}$. The net positive charge increases the source/drain
electron channel conductance. The long-lived electron current
passing through the channel over time is responsible for the
photoconductive gain mechanism.}
\end{figure}

The wavelength dependence of the onset of the single-photon-conductivity
steps is correlated with the 1.9 eV bandgap of the Al$_{0.3}$Ga$_{0.7}$As
layer, on top of the 2D electron gas. The onset of single-photon steps
begins at wavelengths shorter than $\lambda $
\mbox{$<$}%
650 nm, becoming more pronounced at $\lambda $ = 550 nm where the Al$_{0.3}$%
Ga$_{0.7}$As is more absorbing. By contrast, the single-photon steps are not
seen at $\lambda $ = 700 nm. A corresponding model of the photo-conductive
process at $\lambda $ = 550 nm is illustrated in Fig. 4. A photon is
absorbed in the Al$_{0.3}$Ga$_{0.7}$As layer, with the photo-hole being
trapped at DX$^{-}$ centers that are associated with the n-type doping.
Alternately the photo-hole can be trapped at neutral donors, d$^{0}$, though
none are shown in Fig. 4. (According to the negative-U property \cite{DX} of
DX$^{-}$ centers, neutral donors segregate into ionized donors d$^{+}$ and DX%
$^{-}$ centers.) Regardless, in either case, the net increase in positive
charge among the donor defects opens up the source/drain channel creating a
permanent increase in electron current.

In converse, at $\lambda $ = 700 nm, photons are absorbed in the nominally
undoped GaAs buffer layer, which is usually weakly n-type. Thus the
photo-hole recombination centers, residual neutral donors d$^{0}$, are very
dilute. They are too far away from the source/drain channel to produce
noticeable discrete jumps in current for single-photon events at $\lambda $
= 700 nm. Nonetheless, the smooth photoconductance channel quantization
steps are still observed at $\lambda $ = 700 nm, as shown in Fig. 2A.

The photo-electron plays a lesser role. It usually ends up in the channel,
and then becomes swept away in the source and drain electrodes. Since those
are ohmic contacts, they can continue to inject replacement electrons
indefinitely. This is essentially the mechanism \cite{Rose78} of secondary
photoconductivity and photoconductive gain that is responsible for the
single-photon sensitivity.

The conductance curve in Fig. 2A ends at the 2 conductance units (2e$^{2}$%
/h) plateau. Above this conductance level, the gate-induced and
photo-induced conductance changes no longer match. The photo-induced
conductance change tends to saturate above 2 conductance units (2e$^{2}$/h),
after about 70 discrete photon conductance steps, but the gate-induced
change continues to higher conductance. We attribute the saturation of the
photo-induced conductance to saturation of the doping-induced trapping
centers. Within the active area of 3$\times $10$^{-10}$ cm$^{2}$ there are
3.3$\times $10$^{11}$ carriers/cm$^{2}$ or only about 100 carriers,
explaining the saturation in photoconductance. At a channel capacitance of $%
\approx $ 0.1 femto-Farad, the 70 charges produce about the same electric
field as the gate voltage change of $\Delta $V$_{g}$ = 0.2 Volt that was
required to reach the 2 (2e$^{2}$/h) conductance plateau.

In this photoconductive single-photon detector, a photo-hole is trapped,
producing a discrete change in source/drain channel conductance. If the
photosensitive layer were strained, the light/heavy hole degeneracy would be
lifted, and the hole spin degree of freedom might be a viable long-lived
qubit. Strained hole spin coherence has been maintained \cite{Feher60} for
about 100 nsec in p-Silicon. In n-type material however, trapped holes are
subject to electron/hole recombination. It is generally accepted \cite
{Awschalom99} that a trapped photo-electron spin is the preferred qubit
compared to a photo-hole spin. Electron traps may require artificial
engineering, for example, they could be electron potential wells created by
electrostatic gates above a heterointerface.

In addition to changing the sign of the trapped carrier, a further change
may be needed to increase the quantum efficiency. Photoconductive detectors
can inherently be quite efficient, since the photo-carriers are produced by
band-to-band absorption in a direct bandgap semiconductor. Nonetheless, it
might be desirable in practice to incorporate the absorbing region into an
optical cavity to make it a cavity-enhanced photodetector.

In principle, a photoconductive detector can store and detect an optically
injected photo-carrier charge, preserving its quantum mechanical spin
information. This would safely prevent the charge measurement from
disturbing the spin. We have demonstrated single photo-carrier charge
sensitivity, but we have used naturally occurring defect centers that are
subject to carrier recombination. It remains yet to create an artificial
potential well that would safely trap and store the spin.

We thank HongWen Jiang, James Chadi, and Mineo Saito for helpful
discussions. The project is sponsored by the Defense Advanced Research
Projects Agency \& Army Research Office Nos. MDA972-99-1-0017 and
DAAD19-00-1-0172. The content of the information does not necessarily
reflect the position or the policy of the government, and no official
endorsement should be inferred.



\begin{references}
\bibitem{DX}  It is now understood that DX$^{-}$ is a donor ion that has
captured an extra electron, thereby shifting to an interstitial site.\cite
{Chadi88,Chadi89} It has a negative-U property, preferring to take the
electron from another donor in the reaction 2d$^{0}$ $\rightarrow $\ d$^{+}$
+ DX$^{-}$.
\end{references}
\end{document}